\documentclass[12pt,a4paper]{article}
\usepackage{epsfig}
\begin{document}
\title{Neutral top-pion and  the rare top decays $t\rightarrow c l_{i} l_{j} $}
\author{Chongxing Yue, Lei Wang, Dongqi Yu\\
{\small Department of Physics, Liaoning  Normal University, Dalian
116029. P.R.China}
\thanks{E-mail:cxyue@lnnu.edu.cn}}
\date{\today}
\maketitle
\begin{abstract}
\hspace{5mm}We study the rare top decays $t\rightarrow c l_{i}
l_{j}(l=\tau,\mu,or\  e)$ in the framework of topcolor-assisted
technicolor($TC2$) models. We find that the neutral top-pion
$\pi_{t}^{0}$ can produce significant contributions to these
processes via the flavor changing couplings $\pi_{t}^{0} \bar{t}
c$ and $\pi_{t}^{0} l_{i} l_{j}$. For the $\pi_{t}^{0}$ mass
$m_{\pi_{t}}=150 GeV$ and the parameter $\varepsilon=0.08$, the
branching ratio $Br$(t$\rightarrow c \tau \tau )$ can reach
$7.1\times10^{-7}$. Taking into account the constraints of the
present experimental limit of the process $\mu\rightarrow e
\gamma$ on the free parameters of $TC2$ models, we find that the
value of $Br$($t\rightarrow c \tau
\mu$)$\approx$$Br$($t\rightarrow c \tau e$) is in the range of
1.8$\times10^{-10}\sim1.7\times10^{-8}$.
\end {abstract}

\vspace{2.0cm} \noindent
 {\bf PACS number(s)}:14.65.Ha, 14.80.Cp, 12.60.Cn

\newpage
\noindent{\bf I. Introduction}

   It is well known that, in the standard model($SM$), the flavor changing neutral current  \/ $(FCNC)$ is absent at tree-level
   and at one-loop level they are $GIM$ suppressed. The $SM$ results of the
   rare decays, which are induced by $FCNC$, are very small and can
   not be detected in the current or future high energy collider
    experiments. Thus, rare decays provide a very sensitive probe of
    new physics beyond the $SM$. Detection of rare decays at visible
    levels by any of the future colliders would be instant evidence of new physics.
   Searching for rare decays is one of the major goals of the next
   generation of high energy collider experiments.

          The large value of the top quark mass offers the possibility that
     it can be singled out to play a key role in probing the
     new physics beyond the $SM$. The properties of the top quark
     could reveal information regarding flavor physics,
     electroweak symmetry breaking($EWSB$) mechanism, as well as new
     physics beyond the $SM$[1]. One of these consequences is
     that rare top decays can be used to detect new physics. Much
     of the theoretical activity involving rare top decays has
     occurred within some specific models beyond the $SM$[2]. It has
     been shown that the values of the branching ratios  $Br$($t\rightarrow
     cv$)$(v = g, \gamma$, or $z)$ or $Br$($t\rightarrow cvv$)$(v = g, \gamma$, or $w)$
     can indeed be enhanced by
     several orders of magnitude, which may be within the
     observable threshold of near-future-experiments. Recently,
     the rare top decays $t\rightarrow cl_{i}l_{j}(l=\tau,\mu,or\  e)$ have been
     studied in the framework of the general two Higgs doublet
     model[3]. They have shown that the neutral Higgs bosons
     $h^{0}$ and $A^{0}$ can make the branching ratios
     $Br$($t\rightarrow cl_{i}l_{j}$) reach $1 \times 10^{-8}$.

     The presence of the top-pions $\pi_{t}^{0,\pm}$ in low-energy
     spectrum  is an inevitable feature of topcolor scenario[4].
     These new particles have large Yukawa couplings to the
     third family quarks and can induce the tree-level {\em FC}
     couplings, which have significant contributions to the rare
     top decays $t\rightarrow c\gamma(g,z$) and $t\rightarrow cww$[5].
     In this paper, we will study the contributions of the neutral
     top-pion $\pi_{t}^{0}$ to the rare top decays $t\rightarrow
     c l_{i}l_{j}$ in the context of topcolor-assisted
      technicolor($TC2$) models[6]. These rare decay processes can
      occur at the tree-level or one-loop level, which are induced
      by the {\em FC} couplings $\pi_{t}^{0}\bar{t}c$ or $\pi_{t}^{0}
      l_{i}l_{j}(i\neq j)$. Our numerical results show that the neutral
      top-pion $\pi_{t}^{0}$ can generate significant contributions
       to these processes. In all of the parameter space, the values of the branching
      ratio $Br$($t\rightarrow c\tau\tau$) are larger than those of $Br$($t\rightarrow
      c\mu\mu$) or $Br$($t\rightarrow cee$). For $m_{\pi_{t}}=150GeV$
      and the parameter $\varepsilon$=0.08, we have $Br(t\rightarrow
      c\tau\tau)=7.1\times10^{-7}$. Taking into account the
       constraints of the present experimental limit of the process
      $\mu\rightarrow e\gamma$ on the flavor mixing parameter
      $K_{ij}$, we further calculate the contributions of
      $\pi_{t}^{0}$ to the rare top decays $t\rightarrow c\tau\mu$ and $c\tau e$.
      We find that the branching ratio $Br$($t\rightarrow c\tau\mu$) is
      approximately equal to the branching ratio $Br$($t\rightarrow c\tau e$).
      In most of the parameter space, the value of $Br$($t\rightarrow c\tau\mu$)
      is in the range of $1.8\times10^{-10}\sim1.7\times10^{-8}.$

       The paper is organized as follows: In section 2, we will present the
       expressions of the branching ratios $Br$($t\rightarrow cl_{i}l_{j}$)
       contributed by the neutral top-pion $\pi_{t}^{0}$. Our numerical results
       and conclusions are given in section 3.

\noindent{\bf II. The neutral top-pion $\pi_{t}^{0}$ and the
branching ratios $Br$($t\rightarrow cl_{i}l_{j}$)}

For $TC2$ models[4,6], the underlying interactions, topcolor
interactions, are assumed to be chiral critically strong at the
scale about $1TeV$ and coupled preferentially to the third
generation, and therefore do not posses $GIM$ mechanism. The
non-universal topcolor interactions result in the {\em FC}
coupling vertices when one writes the interactions in the mass
eigen basis. Thus, the top-pions can induce the new {\em FC}
scalar coupling vertices. The couplings of the neutral top-pion
$\pi_{t}^{0}$ to ordinary fermions, which are related to the rare
top decays $t\rightarrow cl_{i}l_{j}$, can be written as [7,8]:
\begin{equation}
\frac{m_{t}}{\sqrt{2}F_{t}}\frac{\sqrt{\nu_{w}^{2}-F_{t}^{2}}}{\nu_{w}}
     [K_{UR}^{tt}K_{UL}^{tt^{*}}\bar{t}\gamma^{5}t\pi_{t}^{0}+K_{UR}^{tc}
     K_{UL}^{tt^{*}}\bar{t_{L}}
     c_{R}\pi_{t}^{0}]+\frac{m_{l}}{\sqrt{2}\nu_{w}}\bar{l}
     \gamma^{5}l\pi_{t}^{0}+\frac{m_{\tau}}{\sqrt{2}\nu_{w}}
     K_{\tau i}\bar{\tau}\gamma^{5}l_{i}\pi_{t}^{0}.
\end{equation}
Where $\nu_{w}=\nu / \sqrt{2}=174GeV$ and $F_{t}\approx$
     50$GeV$ is the top-pion decay constant, which can be estimated
     from the Pagels-stokar formula. $K_{UL}$ and $K_{UR}$ are
     rotation matrices that diagonalize the up-quark mass matrix
     $M_{U}$, i.e. $ K_{UL}^{+} M_{U} K_{UR}=M_{U}^{dia}$. To yield a
     realistic form of the $CKM$ matrix $V$, it has been shown that their values
     can be taken as [7]:
      \begin{equation}K_{UL}^{tt}\approx 1,\ \ \  K_{UR}^{tt}=
     1-\varepsilon,\ \ \  K_{UR}^{tc}\leq
     \sqrt{2\varepsilon-\varepsilon^{2}}\ .\end{equation}
In the following calculation, we will take
     $K_{UR}^{tc}=\sqrt{2\varepsilon-\varepsilon^{2}}$ and take
     $\varepsilon$ as a free parameter, which is assumed to be in
     the range of 0.01-0.1[4,6]. $l_{i}(i=1,2)$ is the first
     (second) lepton $e  (\mu)$, $K_{\tau i}$ is the flavor mixing
     factor, which is the free parameter. Certainly, there is also
     the {\em FC} scalar coupling $\pi_{t}^{0}\bar{\mu}e$. However, the
     topcolor interactions only contact with the third generation
     fermions. The flavor mixing between the first and second
     generation fermions is very small, which can be ignored.
     Thus, we have assumed $K_{\mu e}\approx0.$

\begin{figure}[htb]
\vspace{-10cm}
\begin{center}
\epsfig{file=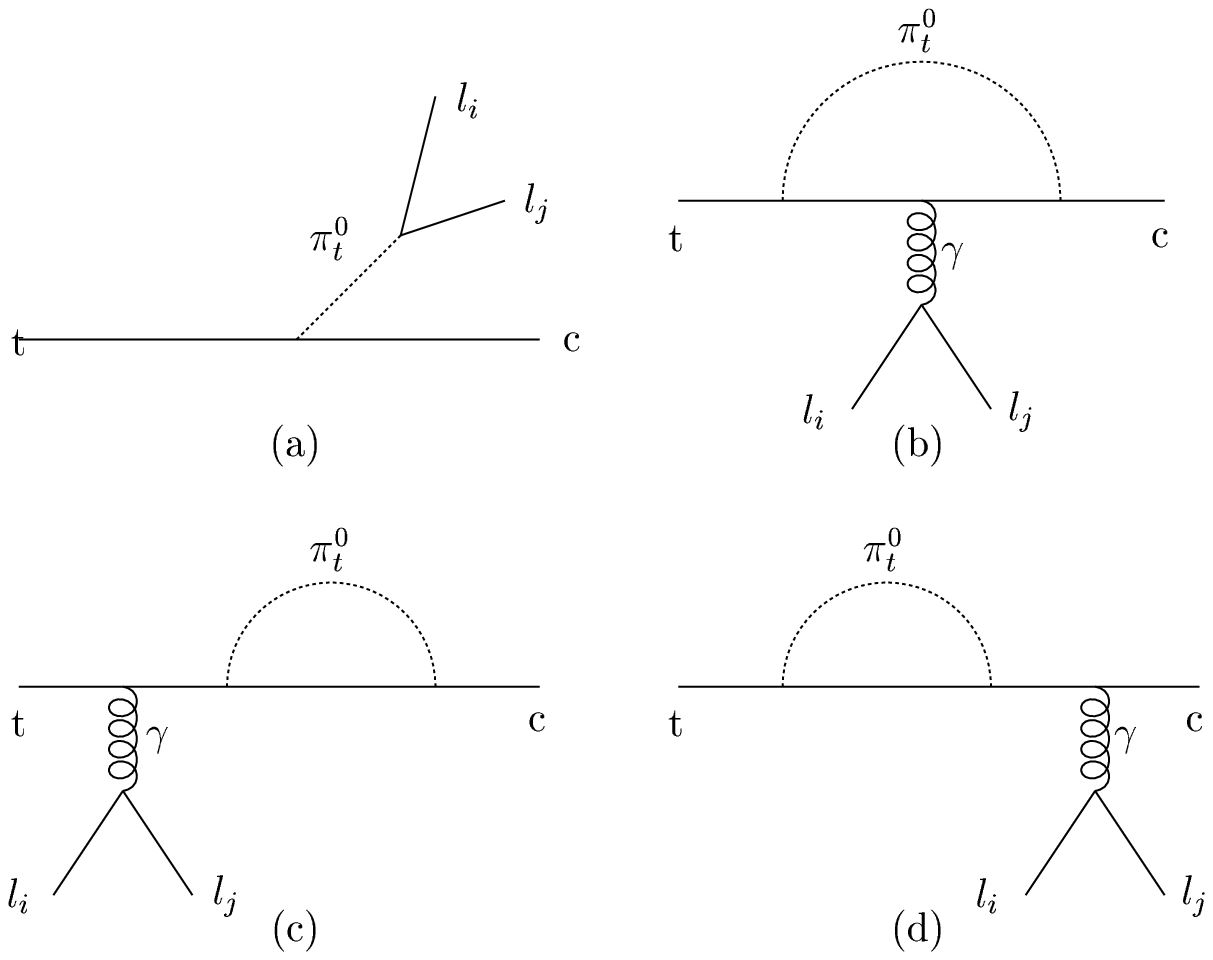,width=550pt,height=800pt} \vspace{-10.0cm}
\hspace{-1cm} \caption{The tree-level and one-loop Feynman
diagrams contribute to the rare top decays \hspace*{1.8cm}
$t\rightarrow cl_{i}l_{j}$ induced by $\pi^{0}_{t}$ exchange in
$TC2$ models.} \label{ee}
\end{center}
\end{figure}

In $TC2$ models, the rare top decays $t\rightarrow cl_{i}l_{j}$
can be induced at the tree-level and can also be induced via
photon penguin diagrams at the one-loop level, as shown in Fig.1.
For the diagrams Fig.1 (b), (c), and (d), we have taken i=j.

Let us first consider the contributions of the neutral top-pion
$\pi_{t}^{0}$ to the rare top decays $t\rightarrow cll
(l=\tau,\mu,or\  e)$. Using Eq.(1), Eq.(2) and other relevant
Feynman rules, the renormalized amplitude can be written as:
\begin{eqnarray}
M_{1}&=&M_{tree}+M_{loop},\\
M_{tree}&=&A_{1}\bar{u}_{c}\gamma_{5}u_{t}\frac{i}{q^{2}-m_{\pi_{t}}^{2}
+im_{\pi_{t}}\Gamma_{total}}\bar{u}_{l}\gamma_{5}v_{l},\\
M_{loop}&=&\bar{u}_{l}(-ie\gamma_{\nu})v_{l}\frac{-ig^{\mu\nu}}
{p_{\gamma}^{2}+i\varepsilon}\bar{u}_{c}\Lambda_{tc\gamma, \mu}u_{t},\\
\Lambda_{tc\gamma, \mu}&=&A_{2}[\gamma_{\mu}F_{1}+p_{t,\mu}
F_{2}+p_{c,\mu}F_{3}]
\end{eqnarray}
with
\begin{eqnarray*}
A_{1}&=&i\frac{m_{t}}{\sqrt{2}F_{t}}\frac{\sqrt{\nu_{w}^{2}-F_{t}^{2}}}
{\nu_{w}}K_{UR}^{tc}K_{UL}^{tt^{*}}\frac{m_{l}}{\nu},
\hspace{0.5cm}
A_{2}=\frac{i}{16\pi^{2}}\frac{em_{t}^{2}}{3F_{t}^{2}}
\frac{\nu_{w}^{2}-F_{t}^{2}}{\nu_{w}^{2}}K_{UR}^{tc}K_{UR}^{tt}
(K_{UL}^{tt^{*}})^{2}
\end{eqnarray*}
\begin{eqnarray*}
F_{1}&=&[(m_{c}-m_{t})(-m_{t}C_{11}+m_{t}C_{12}+m_{c}C_{12})
-2C_{24}+m_{\pi}^{2}C_{0}+B_{0}-B_{0}^{*}
-B_{1}^{'}],\nonumber\\
F_{2}&=&2m_{c}(C_{22}-C_{23})+2m_{t}(2C_{23}-C_{21}-C_{22}),\nonumber\\
F_{3}&=&-2(m_{c}-m_{t})(C_{12}+C_{22})-2m_{t}C_{23}.\nonumber\
\end{eqnarray*}
Where $q$ is the four momenta of the neutral top pion
$\pi_{t}^{0}$ and $\Gamma_{total}$ is the total decay width of
$\pi_{t}^{0}$, which has been calculated in Ref.[9]. The
expressions of the two-and three-point scalar integrals $B_{n}$
and $ C_{ij}$ are [10]:
\begin{eqnarray*}
B_{n}&=&B_{n}(-\sqrt{\hat{s}},m_{t},m_{t}),\\
B^{*}_{n}&=&B_{n}(-P_{c},m_{\pi_{t}},m_{t}),\\
B_{n}'&=&B_{n}(-P_{t},m_{\pi_{t}},m_{t}),\\
C_{ij}&=&C_{ij}(p_{t},-\sqrt{\hat{s}},m_{\pi_{t}},m_{t},m_{t}),\\
C_{0}&=&C_{0}(p_{t},-\sqrt{\hat{s}},m_{\pi_{t}},m_{t},m_{t}).
\end{eqnarray*}

For the rare top decay processes $t\rightarrow c\tau l
 \ (l=\mu\  or\  e )$, the neutral top-pion $\pi_{t}^{0}$ can only
have contribution to these processes via Fig.1 (a). The
renormalized amplitude can be written as:
\begin{eqnarray}
M_{2}&=&A_{3}\bar{u}_{c}\gamma_{5}u_{t}\frac{i}{q^{2}-m_{\pi_{t}}^{2}+
im_{\pi_{t}}\Gamma_{total}}\bar{u}_{\mu}\gamma^{5}v_{\tau},\\
A_{3}&=&i\frac{m_{t}}{\sqrt{2}F_{t}}\frac{\sqrt{\nu_{w}^{2}-
F_{t}^{2}}}{\nu_{w}}K_{UR}^{tc}K_{UL}^{tt^{*}}\frac{m_{\tau}}{\nu}K_{\tau
l}\ .
\end{eqnarray}

The decay width $\Gamma(t\rightarrow cl_{i}l_{j})$ can be obtained
in the usual way and it is given by
\begin{eqnarray}
\Gamma(t\rightarrow cl_{i}l_{j})=\frac{1}{2m_{t}}\int(2\pi)^{4}
\delta^{4}(p_{t}-p_{c}-p_{l_{i}}-p_{l_{j}})
\bar{\Sigma}|M|^{2}\frac{d^{3}p_{c}}{(2\pi)^{3}2E_{c}}
 \frac{d^{3}p_{l_{i}}}{(2\pi)^{3}2E_{l_{i}}}
 \frac{d^{3}p_{l_{j}}}{(2\pi)^{3}2E_{l_{j}}},
\end{eqnarray}
 where $E_{c},$ $E_{l_{i}},$ and $E_{l_{j}}$ are the energies of the
 final particles $ c$, $l_{i},$ and $l_{j}$, respectively. In the
 following section, we will use these formulae to give our numerical
 results.
\begin{figure}[htb]
\vspace{-1cm}
\begin{center}
\epsfig{file=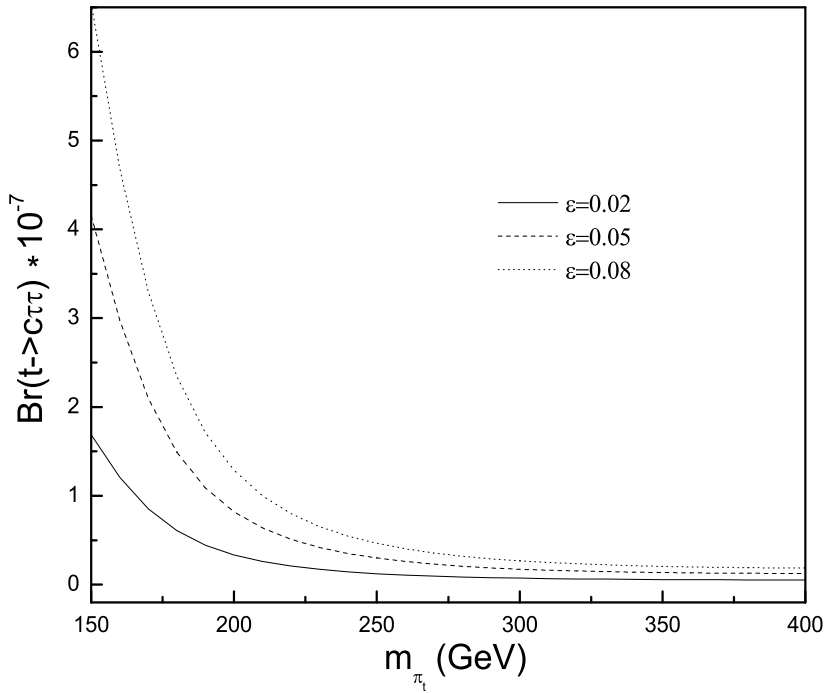,width=350pt,height=300pt} \vspace{-1cm}
\hspace{1cm} \caption{The branching ratio $Br(t\rightarrow
c\tau\tau)$ as a function of the top-pion mass $m_{\pi_{t}}$ for
\hspace*{1.8cm}three values of the free parameter $\varepsilon$,
in which we have included the tree-level \hspace*{1.8cm}and
one-loop contributions of $\pi_{t}^{0}$ exchange.} \label{ee}
\end{center}
\end{figure}

\noindent{\bf III. Numerical results and conclusions}

To calculate the branching ratios of the rare top decays
$t\rightarrow cl_{i}l_{j}$, we assume that the total decay width
$\Gamma_{t}$ of the top quark is dominated by the decay channel
$t\rightarrow wb$, which has been taken $\Gamma (t\rightarrow
wb)=1.56GeV$. In our calculation, we have taken $m_{t}=175GeV$,
$m_{c}=1.2GeV$, $m_{\tau}=1.78GeV$, $m_{\mu}=0.105GeV$, and
$\alpha_{e}=\frac{1}{128.8}$[11]. Except for these input
parameters, there are three free parameters in the expressions of
the branching ratios $Br$($t\rightarrow cl_{i}l_{j}$):
$\varepsilon, m_{\pi_{t}}$ and
 $K_{\tau l}$. To avoid significant fine-tuning, Ref.[6] has shown
that the free parameter $\varepsilon$ should be in the range of
$0.01\sim0.1$. The limits on the mass $m_{\pi_{t}}$ of the
top-pion may be obtained via studying it$^{,}$s effects on various
experimental observables [4,12]. It has been shown that the
$m_{\pi_{t}}$ is allowed to be in the range of a few hundred $GeV$
depending on the models. As numerical estimation, we take the
$\pi_{t}^{0}$ mass $m_{\pi_{t}}$ to vary from 150$GeV$ to
400$GeV$. Our numerical results are summarized in Fig.2$\sim$
Fig.4.

\begin{figure}[htb]
\vspace{-0.5cm}
\begin{center}
\epsfig{file=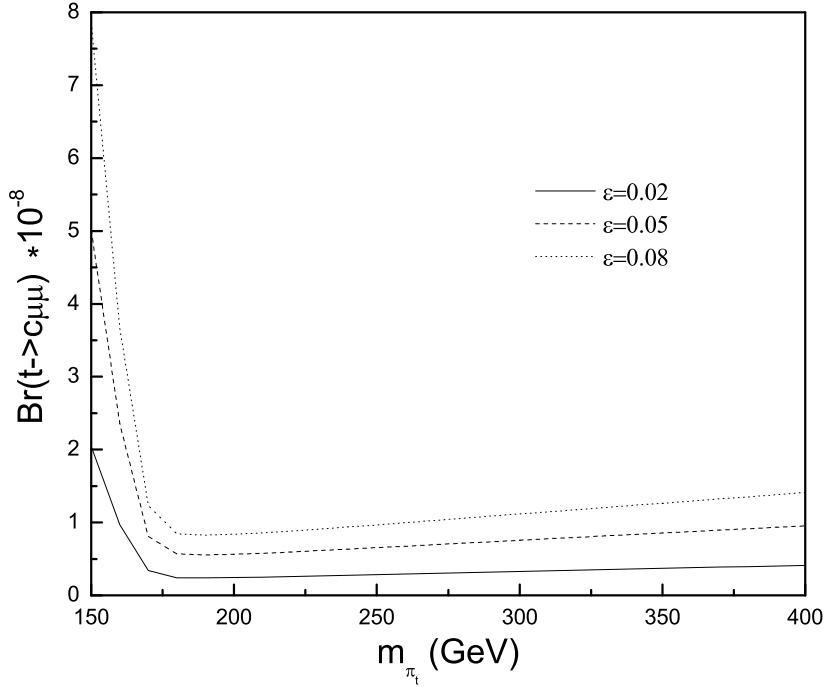,width=350pt,height=300pt} \vspace{-1.0cm}
\hspace{1cm} \caption{Same as Fig.2 but for the rare top decay
$t\rightarrow c\mu\mu$.} \label{ee}
\end{center}
\end{figure}

The branching ratios $Br$($t\rightarrow c\tau\tau$) and
$Br$($t\rightarrow c\mu\mu$) are plotted as functions of the
$\pi_{t}^{0}$ mass $m_{\pi_{t}}$ for three values of the free
parameter $\varepsilon$ in Fig.2 and Fig.3, respectively. From
these figures, we can see that the values of these branching
ratios increase as $\varepsilon$ increasing. In all of the
parameter space of the $TC2$ models, the values of branching ratio
$Br$($t\rightarrow c\tau\tau$) are larger than those of
$Br$($t\rightarrow c\mu\mu$). For example, for
$m_{\pi_{t}}=200GeV$ and $\varepsilon=0.05$, the values of
$Br$($t\rightarrow c\tau\tau$) and $Br$($t\rightarrow c\mu\mu$)
 are $8.2\times10^{-8}$ and $5.6\times10^{-9}$, respectively.
 This is because the one-loop contributions of
  $\pi_{t}^{0}$ exchange to the branching ratios $Br(t\rightarrow
  c\tau\tau)$, $Br(t\rightarrow c\mu\mu)$, and $Br(t\rightarrow cee)$
 are approximately equal to each other, which are smaller than
 $5\times10^{-9}$. However, the tree-level contributions of $\pi_{t}^{0}$ exchange
 to the branching ratios $Br$($t\rightarrow cll$) are proportional to
 $\frac{m_{l}^{2}}{\nu}$. Thus, for the process $t\rightarrow
 c\tau\tau$, the tree-level contributions are lager than the one-loop
 contributions at least by two orders of magnitude. For the process
 $t\rightarrow c\mu\mu$, two kinds of the
 contributions are comparable.

 For the process $t\rightarrow cee$,
 the factor $\frac{m_{e}^{2}}{\nu^{2}}$ strongly suppresses the
 tree-level contributions, the values of $Br(t\rightarrow cee)$
 is mainly generated by the one-loop contributions, which is
 smaller than $5\times10^{-9}$. Thus, if we would like to detect
 the possible signals of the neutral top-pion $\pi_{t}^{0}$ at the
 future high energy experiments via the rare top decays
 $t\rightarrow cll$, we should first consider the process $t\rightarrow
 c\tau\tau$. For instance, if we assume $m_{\pi_{t}}=150GeV$ and
 $\varepsilon=0.08$, then the values of the branching ratio
 $Br$($t\rightarrow c\tau\tau$) can reach $7.1\times10^{-7}$.

\begin{figure}[htb]
\vspace{-0.5cm}
\begin{center}
\epsfig{file=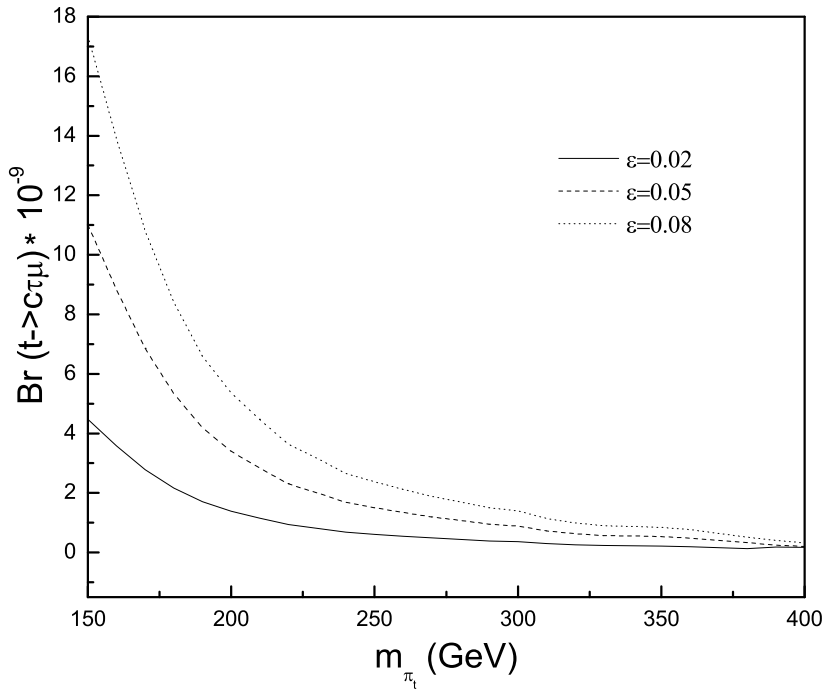,width=350pt,height=300pt} \vspace{-1.0cm}
\hspace{1cm} \caption{The branching ratio $Br(t\rightarrow
c\tau\mu)\approx Br(t\rightarrow c\tau e)$ as a function of the
$\pi_{t}^{0}$ mass \hspace*{1.8cm}$m_{\pi_{t}}$ for three values
of the free parameter $\varepsilon$.}  \label{ee}
\end{center}
\end{figure}

 From above discussions, we can see that the neutral top-pion
 $\pi_{t}^{0}$ can only produce the tree-level contributions to
 the rare top decays $t\rightarrow c\tau l(l=\mu\  $or $e$) via the
 Feynman diagram Fig.1(a). The expressions of the branching ratios
 $Br$($t\rightarrow c\tau l$) are dependent on the $\pi_{t}^{0}$
 mass $m_{\pi_{t}}$, the flavor mixing factor $K_{\tau l}$, and the
 parameter $\varepsilon$. Furthermore, $\pi_{t}^{0}$ can produce
 significant contributions to the lepton flavor violating($LFV$)
 processes $l_{i}\rightarrow l_{j}\gamma$ via the {\em FC} scalar
 couplings $\pi_{t}^{0}\tau l$. For the $LFV$ process
  $\mu\rightarrow e\gamma$, the neutral top-pion  $\pi_{t}^{0}$ can
 contribute this process via the on-shell photon penguin diagrams
 generated by the FC scalar coupling $\pi_{t}^{0}\tau \mu$ and
 $\pi_{t}^{0}\tau e$. Thus, using the present
 experimental bound on the $LFV$ process $\mu\rightarrow e\gamma$,
 i.e. $Br$($\mu\rightarrow e\gamma)\leq1.1\times10^{-11}$, we can
 give the constraints on the flavor mixing factor $ K $ for
 $150GeV\leq m_{\pi_{t}}\leq 400GeV$, in which we have assumed $K=K_{\tau
 \mu}=K_{\tau e}$. If we take that $m_{\pi_{t}}$ is smaller than
 $400GeV$, than there must be $K \leq0.2$[8]. Taking into account the
 constraints on the flavor mixing factor $ K $, the branching ratio
 $Br$($t\rightarrow c\tau \mu$)$\approx$$Br$($t\rightarrow c\tau e$) is
 plotted as a function of $m_{\pi_{t}}$ for three values of the
 parameter $\varepsilon$ in Fig.4. One can see, from Fig.4, that
 the value of $Br$($t\rightarrow c\tau\mu$) increases as
 $m_{\pi_{t}}$ decreasing and $\varepsilon$ increasing. For
 $0.02\leq\varepsilon\leq0.08$ and $150GeV\leq m_{\pi_{t}}\leq
 400GeV$, the value of $Br$($t\rightarrow c\tau\mu$)$\approx
 Br(t\rightarrow c\tau e)$ is in the range of $1.8\times10^{-10}
 \sim1.7\times10^{-8}$.

 The $CERN$ $LHC$ will allow to probe the couplings of  the top
 quark to both known and new particles involved in possible top
 decay channels different from the main $t\rightarrow wb$. It is
 very useful to analyze the rare top decays for detecting the new physics beyond the $SM$[1,2].
  If we assume that the production cross section of the top quark pairs is about
 $8\times10^{5}fb$ at the $LHC$ with a yearly integrated
 luminosity of $\pounds=100fb^{-1}$, then we can only obtain 8 $c\tau\tau$
 events per year for $m_{\pi_{t}}=150GeV$ and $\varepsilon=0.08$.

 The signal of the rare top decay $t\rightarrow c\tau\tau$ is the
 isolated $ \tau $ leptons with a $ c$-quark jet. If we want to
 detect this signal, we must consider the efficiency problems in
 measuring the $ \tau $ lepton and identifying a $ c$-quark jet.
 Furthermore, we will have to take the suitable kinematical cuts
 to extract the signal from a possible large reducible background,
 which will further degrade the signal. Thus, it is very difficult
 to detect the possible signals of  the neutral top-pion
 $\pi_{t}^{0}$ by the process $t\rightarrow c\tau\tau$ at the future $LHC$
 experiments. However, if the topcolor scenario or other new physics
 beyond the SM can predict the existence of a light neutral
 scalar, for example $m_{s}<100GeV$, which can induce the $FC$ scalar
 couplings, its signals might be detected via the process
 $t\rightarrow c\tau\tau$ at the future $LHC$ experiments. Even if
 we can not observe the rare top decay $t\rightarrow c\tau\tau$
induced by the light neutral scalar, we can obtain experimental
bounds on the masses of this kind of new particles[14].

 The possibility of detecting the top-pions via the $FC$ processes
 at the $LHC$ experiments has been extensively studied in
 Ref.[7,13]. They have shown that the signals of top-pions
 should be observable at the $LHC$ in a sizable region of the
 parameter space. Therefore, we have to say that the top-pions are
 more easy detected using these $FC$ processes than using the rare
 top decays $t\rightarrow cl_{i}l_{j}$. However, our work provides
 one possible method to study the signals of the neutral scalar
 at the $LHC$ experiments.

 \vspace{0.5cm} \noindent{\bf Acknowledgments}

Chongxing Yue would like to thank the {\bf Abdus Salam }
International Centre for Theoretical Physics(ICTP) for partial
support. This work was supported in part by the National Nature
Science Foundation of China (90203005) and the Natural Science
Foundation of the Liaoning Scientific Committee(20032101).

\end{document}